\documentclass[letter]{aa} 
\usepackage{graphicx}
\usepackage{txfonts}
\usepackage{lipsum}
\usepackage{lastpage}
\usepackage{xpatch}
\usepackage{hyperref}
\hypersetup{
	colorlinks=true,
	breaklinks=true,
	citecolor=blue,
	allcolors=blue,
	frenchlinks=true
}

\makeatletter
\xpatchcmd\NAT@citex
{%
	\@citea\NAT@hyper@{%
		\NAT@nmfmt{\NAT@nm}%
		\hyper@natlinkbreak{\NAT@aysep\NAT@spacechar}{\@citeb\@extra@b@citeb}%
		\NAT@date
	}%
}
{%
	\@citea
	\NAT@nmfmt{\NAT@nm}%
	\NAT@aysep\NAT@spacechar
	\NAT@hyper@{\NAT@date}%
}
{}{}
\xpatchcmd\NAT@citex
{%
	\@citea\NAT@hyper@{%
		\NAT@nmfmt{\NAT@nm}%
		\hyper@natlinkbreak{\NAT@spacechar\NAT@@open\if*#1*\else#1\NAT@spacechar\fi}%
		{\@citeb\@extra@b@citeb}%
		\NAT@date
	}%
}
{
	\@citea
	\NAT@nmfmt{\NAT@nm}%
	\NAT@spacechar\NAT@@open\if*#1*\else#1\NAT@spacechar\fi
	\NAT@hyper@{\NAT@date}%
}
{}{}
\makeatother

\makeatletter
\renewcommand*\aa@pageof{, page \thepage{} of \pageref*{LastPage}}
\makeatother

\begin{document}

   \title{How supermassive black holes shape central entropies in galaxy clusters}

   \author{R. Weinberger\thanks{rweinberger@aip.de} 
        \and C. Pfrommer
        }

\institute{Leibniz Institute for Astrophysics Potsdam (AIP),
              An der Sternwarte 16, 14482 Potsdam, Germany }

   \date{Received \today}

  \abstract{A significant fraction of galaxy clusters show central cooling times of less than 1~Gyr and associated central cluster entropies below $30\,\mathrm{keV}\,\mathrm{cm}^2$. We provide a straight forward explanation for these low central entropies in cool core systems and how this is related to accretion onto supermassive black holes (SMBHs). Assuming a time-averaged equilibrium between active galactic nucleus (AGN) jet heating of the radiatively cooling intracluster medium (ICM) as well as Bondi accretion, we derive an equilibrium entropy that scales with the SMBH and cluster mass as $K\propto M_\bullet^{4/3}M_{500\mathrm{c}}^{-1}$. At fixed cluster mass, overly massive SMBHs would raise the central entropy above the cool core threshold, thus implying a novel way of limiting SMBH masses in cool core clusters. We find a limiting mass of $1.4\times10^{10}\,\mathrm{M}_\odot$ in a cool core cluster of mass $10^{15}\,\mathrm{M}_\odot$. We carry out three-dimensional hydrodynamical simulations of an idealized Perseus-like cluster with AGN jets and find that they reproduce the predictions of our analytic model, once corrections for elevated jet entropies are applied in calculating X-ray emissivity-weighted cluster entropies. Our findings have significant implications for modelling galaxy clusters in cosmological simulations: a combination of overmassive SMBHs and high heating efficiencies preclude the formation of cool core clusters. }

   \keywords{Galaxies: clusters: intracluster medium -- 
                Galaxies: jets --
                Methods: analytical -- 
                Methods: numerical
               }

   \maketitle

   \nolinenumbers

\section{Introduction}

Galaxy clusters exhibit increasing entropy profiles with radius in the outskirts that are predominantly caused by cosmological gas accretion onto clusters \citep{Voit2005}. At smaller radii, the entropy profiles display a large diversity and show a bimodal distribution in core entropy values \citep{Cavagnolo2009,Hudson2010} which are separated by a critical cluster electron entropy of 
\begin{align}
    K_\mathrm{e,\,crit}=\left.\frac{k_\mathrm{B} T_\mathrm{e}}{n_\mathrm{e}^{2/3}}\right|_\mathrm{crit}=30~\mathrm{keV~cm}^{2},
\end{align}
where $k_\mathrm{B}$ denotes Boltzmann's constant,  and $T_\mathrm{e}$ and $n_\mathrm{e}$ denote the electron temperature and density, respectively. Clusters with central entropy values exceeding $K_\mathrm{e,\,crit}$ are non-cool core systems with comparatively hot cores that are characterized by large central gas cooling times ($\tau_\mathrm{cool}>1$~Gyr). By contrast, clusters with central entropy values below $K_\mathrm{e,\,crit}$ show a temperature dip towards the centre \citep{Vikhlinin2006} and have short central gas cooling times ($\tau_\mathrm{cool}<1$~Gyr).

In the absence of heating, the cores of cool-core clusters would be expected to cool rapidly and sustain star formation at rates of up to several hundred $\mathrm{M}_\odot~\mathrm{yr}^{-1}$ \citep[see][for a review]{Peterson2006}. Observationally, however, both gas cooling and star formation occur at levels far below those predicted by classical, unimpeded cooling-flow models. High-resolution observations with {\it Chandra} and {\it XMM-Newton} reveal a smooth decline in gas temperature toward cluster centres, yet show little evidence for emission from gas cooler than $\sim$1 keV \citep[see the multi-temperature models of][]{Hudson2010}. This suggests that radiative cooling is balanced by a heating mechanism, most likely linked to feedback from the AGN, as indicated by jet-inflated radio lobes spatially coincident with X-ray cavities and a positive correlation of lobe enthalpy and central ICM entropy \citep[Fig.~6 of][]{Pfrommer2012}, which demonstrates that a single AGN outburst is unable to transform a cool core to a non-cool core cluster on the buoyant rise time. Note that part of the observed deficit of soft X-ray emission may be attributable to X-ray absorption \citep{Fabian2022}. Nevertheless, the coupled evolution of gas cooling, star formation, and nuclear activity therefore appears to operate within a self-regulated feedback loop \citep{Voit2005b, McNamara2007, McNamara2012, Fabian2012}. Understanding what sets the central entropy structure in galaxy clusters is a key part of understanding the cooling-flow problem \citep{Peterson2003}.

Simulations of galaxy clusters have long struggled to reliably predict the central entropy profiles. Initial attempts suffered from different numerical methodologies producing unphysical central entropy profiles in non-radiative simulations (steep power-law profiles in smoothed-particle hydrodynamics and too extended cores with adaptive mesh-refinement, \citealt{Frenk1999, Voit2005}), overshadowing possible physical mechanisms causing entropy cores. The inclusion of radiative cooling, AGN heating and cosmic rays \citep[e.g.][]{Sijacki2006, Cattaneo2007,Sijacki2008,Ehlert2018} added to the complexity. The modelling of AGN feedback in particular added a significant degree of freedom to numerical models, creating difficulties in precisely attributing reported differences. To this day, numerical models struggle to reliably reproduce the central entropy cores of groups and clusters \citep[e.g.][]{Barnes2017, Barnes2018, Altamura2023, Pakmor2023, Hernandez-Martinez2025}, with some simulations reporting cool-core clusters emerging in cluster below $10^{15}\,\mathrm{M}_\odot$ but not above \citep[Fig.~8 of][]{Lehle2024}. Importantly, no model fully reproduces the observed central entropy statistics. 

In light of these mixed simulation results, we revisit the problem of central entropy profiles in galaxy clusters from an analytic perspective \citep[see e.g.][]{Nulsen2004, Binney2005, Voit2005b, Voit2015}, with a particular focus on the behaviour of hydrodynamical simulations. To this end, we use an isolated galaxy cluster with a self-regulated AGN jet heating model (\citealt{Ehlert2023,Weinberger2023}, advancing earlier work by \citealt{Weinberger2017b}) to test our analytic considerations.
The letter is structured as follows: in Section~\ref{sec:model} we derive the central entropy of a cool-core galaxy clusters from a set of simple assumptions.
We test this model using hydrodynamical simulations in Section~\ref{sec:sim} and conclude in Section~\ref{sec:conclusion}.

\section{Analytic model}
\label{sec:model}

Consider a cool-core cluster with a SMBH in its centre. We assume the accretion rate $\dot{M}$ to follow the Bondi--Hoyle--Lyttleton formula \citep{Hoyle1939,Bondi1944,Bondi1952,Edgar2004}:
\begin{align}
    \dot{M} &= \frac{4 \pi G^2 M_\bullet^2 \rho}{(\varv^2 + c_\mathrm{s}^2)^{3/2}} 
    \approx \frac{4 \pi G^2 M_\bullet^2}{(\gamma\, K)^{3/2} },
\end{align}
where $G$ denotes the gravitational constant, $M_\bullet$ is the SMBH mass, $P$ and $\rho$ denote the surrounding pressure and density, and  $c_\mathrm{s}=\sqrt{\gamma P \rho^{-1}}$ and $K=P\,\rho^{-\gamma}$ are the sound speed and pseudo-entropy, respectively, where $\gamma=5/3$ is the adiabatic index of the gas. In the last step, we assumed that the SMBH has a negligible velocity in the cluster potential minimum in comparison to the sound speed: $\varv \ll c_\mathrm{s}$. Observationally, the cluster electron entropy,
\begin{align}
    K_\mathrm{e} &\equiv \frac{k_\mathrm{B}\,T_\mathrm{e}}{n_\mathrm{e}^{2/3}} = 
    \frac{\mu m_\mathrm{p}^{5/3}}{(X_\mathrm{e}\,X_\mathrm{H})^{2/3}} K,
\end{align}
is a directly accessible quantity. Here, we assume that the electron temperature $T_\mathrm{e}$ equals the ion temperature $T$, an electron abundance of $X_\mathrm{e} = n_\mathrm{e} / n_\mathrm{H}=1.16$, a mean atomic weight $\mu=0.6$ for a fully ionized primordial gas with hydrogen mass fraction $X_\mathrm{H}=0.76$, and $m_\mathrm{p}$ denotes the proton mass. The accretion rate can therefore be written as 
\begin{align}
    \dot{M} &= \frac{4 \pi G^2 M_\bullet^2 \mu^{3/2} m_\mathrm{p}^{5/2}}{(\gamma\, K_\mathrm{e})^{3/2} X_\mathrm{e} X_\mathrm{H}} \\
    &\approx 3.0\times10^{-2} \,\mathrm{M}_\odot\,\mathrm{yr}^{-1} \, \left(\frac{M_\bullet}{10^{10}\,\mathrm{M}_\odot}\right)^2 \, \left(\frac{K_\mathrm{e}}{30\,\mathrm{keV}\,\mathrm{cm}^2}\right)^{-3/2}.
\end{align}

A fraction $\epsilon$ of the accreted rest mass energy is accelerated as a collimated jet out to kpc scales and heats the surrounding ICM in the cluster core \citep[see e.g.][for a discussion on the entropy increase due to an AGN jet outburst]{Voit2005b}, yielding a heating rate of
\begin{align}
    \dot{E} &= \epsilon \dot{M} c^2 \\
    &\approx 3.4 \times 10^{44}\,\mathrm{erg}\,\mathrm{s}^{-1} \, \left(\frac{\epsilon}{0.2}\right) \, \left(\frac{M_\bullet}{10^{10}\,\mathrm{M}_\odot}\right)^2 \, \left(\frac{K_\mathrm{e}}{30\,\mathrm{keV}\,\mathrm{cm}^2}\right)^{-3/2},
\end{align}
where $c$ is the vacuum speed of light. We can compute the equilibrium central entropy by assuming that the energy injection from the jets matches the cooling luminosity if averaged over timescales longer than the jet duty cycle, $L_\mathrm{cool} = \dot{E}$ \citep{Birzan2004, Dunn2006, Hlavacek-Larrondo2015}:
\begin{align}
    K_\mathrm{e} &= 
    \frac{\mu m_\mathrm{p}^{5/3}}{\gamma}
    \left(\frac{4\pi G^2 \epsilon M_\bullet^2 c^2}{X_\mathrm{e} X_\mathrm{H} L_\mathrm{cool}
    }\right)^{2/3}\\
    &\approx 14.5 \,\mathrm{keV}\,\mathrm{cm}^2 \, \left(\frac{\epsilon}{0.2}\right)^{2/3} \left(\frac{M_\bullet}{10^{10}\,\mathrm{M}_\odot}\right)^{4/3}\,\left(\frac{L_\mathrm{cool}}{10^{45}\,\mathrm{erg}\,\mathrm{s}^{-1}}\right)^{-2/3}. \label{eq:model_lum}
\end{align}
Identifying the ICM cooling luminosity with the bolometric X-ray luminosity, $L_\mathrm{cool}=L_\mathrm{bol}$, we can express $L_\mathrm{bol}$ in terms of a characteristic cluster mass by means of observed galaxy cluster scaling relations (using the relation by \citealt{Chiu2022}, which is consistent with \citealt{Mantz2010}):
\begin{align}
    L_\mathrm{bol} \approx 7\times 10^{44}\,\mathrm{erg}\,\mathrm{s}^{-1} \left(\frac{M_{500\mathrm{c}}}{10^{15}\,\mathrm{M}_\odot}\right)^{3/2}
\end{align}
so that we can write the cluster electron entropy in terms of SMBH mass and cluster mass:
\begin{align}
    K_\mathrm{e} &\approx 18.3 \,\mathrm{keV}\,\mathrm{cm}^2 \, \left(\frac{\epsilon}{0.2}\right)^{2/3} \, \left(\frac{M_\bullet}{10^{10}\,\mathrm{M}_\odot}\right)^{4/3} \, \left(\frac{M_{500\mathrm{c}}}{10^{15}\,\mathrm{M}_\odot}\right)^{-1}. 
\end{align}
This implies a limiting SMBH mass of $1.4\times10^{10}\,\mathrm{M}_\odot$ in a cool core cluster of mass $10^{15}\,\mathrm{M}_\odot$ (assuming $\epsilon=0.2$, adopted from the kinetic feedback efficiency in \citealt{Weinberger2017}). Conversely, for a fixed SMBH mass, the core entropy increases towards smaller galaxy clusters. However, the central galaxies in these systems are found to host significantly less massive SMBHs, which may overcompensate for the decrease in halo mass \citep{Booth2010}. This highlights the need for a deeper understanding of the relationship between SMBH and halo masses in this regime.

\section{Comparison to simulations}
\label{sec:sim}

\begin{figure}
    \centering
    \includegraphics[width=1.0\linewidth]{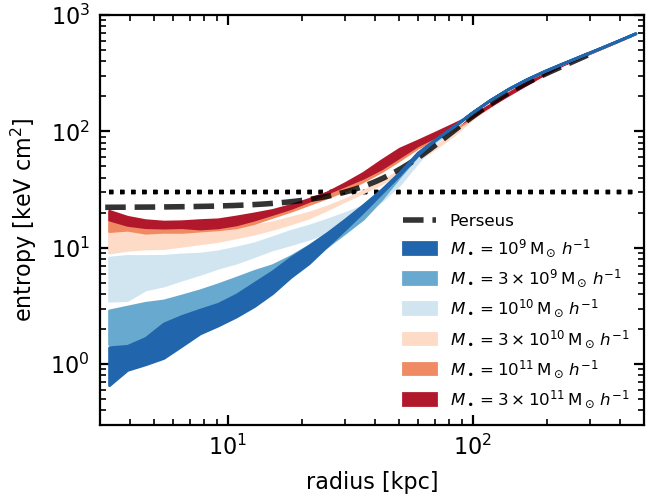}
    \caption{Entropy profiles of X-ray emitting gas of Perseus analogue self-regulated simulations with different SMBH masses (see legend). The coloured area indicates the 16th to 84th percentile of snapshots at times between $t=0.55\,\mathrm{Gyr}$ and $t=1\,\mathrm{Gyr}$. The dotted line denotes $30$~keV~cm$^{2}$, an empirical dividing line between cool-core and non-cool-core clusters. The dashed line denotes the entropy profile of the Perseus cluster from \citet{Churazov2003}, rescaled to $h=0.67$. The central entropy establishes an equilibrium value dependent on SMBH mass.}
    \label{fig:entropyprofile}
\end{figure}

\begin{figure}
    \centering
    \includegraphics[width=1.0\linewidth]{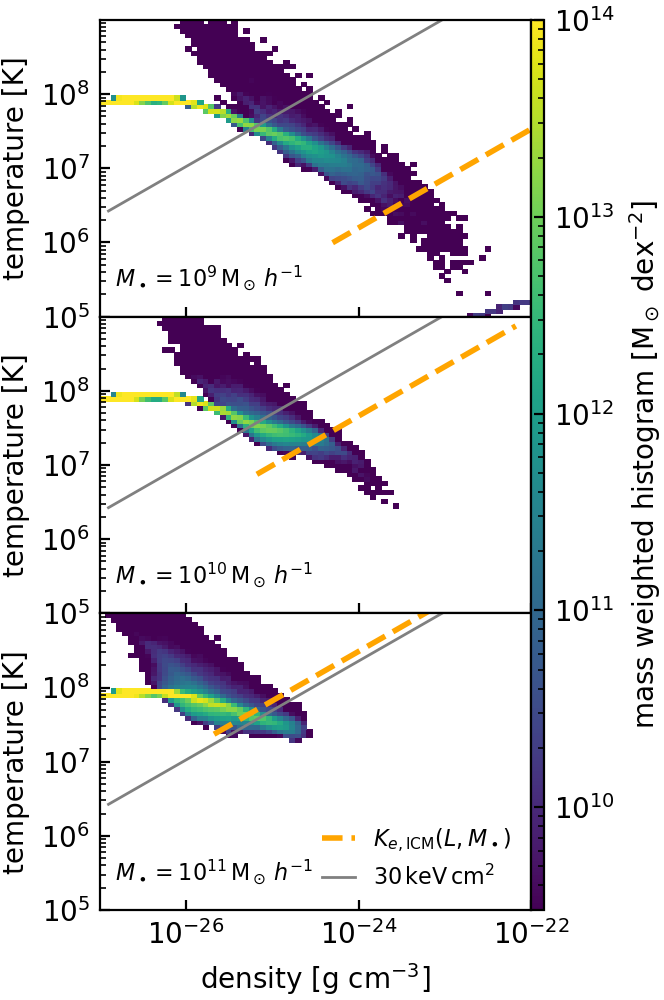}
    \caption{Temperature vs. density phase diagram, colour coded by mass. The grey line indicates an electron entropy of $30\,\mathrm{keV}\,\mathrm{cm}^2$, the orange dashed line indicates the equilibrium ICM entropy predicted by our model (see text for details). Most of the gas mass is above our equilibrium entropy value, implying that efficient AGN jet feedback sets a SMBH mass-dependent entropy floor in the central ICM.}
    \label{fig:phasediagram}
\end{figure}

To test if our model holds in a more realistic setting, we run three-dimensional hydrodynamical simulations as presented in \citet{Weinberger2023}, using the isolated Perseus cluster analogue initial conditions presented in \citet{Ehlert2023}. We restrict ourselves to simulations with slightly lower resolution in the ICM than the fiducial model in \citet{Ehlert2023}, i.e., a target gas mass resolution of $10^7\,\mathrm{M}_\odot$ because this is sufficient to show the steady state behaviour and we aim to match achievable resolutions in cosmological zoom simulations of galaxy clusters. The spatial resolution in the jet is kept identical to the fiducial resolution of \citet{Ehlert2023} at $0.65$~kpc.  While there are some differences in the cold gas properties between magneto-hydrodynamic and purely gas-dynamical models, both simulations self-regulate to a similar ICM \citep{Ehlert2023}. 

To study the scalings predicted in the central entropy model, we ran simulations with different SMBH masses, from $M_\bullet = 10^9 \,\mathrm{M}_\odot\,h^{-1}$ to $M_\bullet = 3\times10^{11}\,\mathrm{M}_\odot\,h^{-1}$, using $h=0.67$. Since efficiency $\epsilon$ and SMBH mass squared $M_\bullet^2$ are degenerate (the SMBH growth through accretion is negligible in this setup) we fixed the efficiency at $\epsilon = 0.2$. We let the simulation equilibrate for $0.55\,\mathrm{Gyr}$ and analyse the $5$ snapshots between $0.55\,\mathrm{Gyr}$ and $1\,\mathrm{Gyr}$ to study the dynamical equilibrium and its scatter.

Figure~\ref{fig:entropyprofile} shows the entropy profile of X-ray emitting gas ($0.2\,\mathrm{keV} < k_\mathrm{B} T < 10\,\mathrm{keV}$), calculated using the emission-weighted electron number density and the emission weighted temperature profiles. The coloured regions show the $16$th to $84$th percentiles of the entropy values at different times. The electron entropy at small radii clearly depends on the SMBH mass, with lower mass SMBHs leading to lower central entropies and more massive SMBHs equilibrating at higher entropy values. Quantitatively, however, the SMBH mass is varied by a factor of 300, which, according to Eq.~\eqref{eq:model_lum}, would imply an equilibrium entropy change of roughly 2000. In contrast, the X-ray emissivity-weighted entropy differs by only about 30. This discrepancy arises from the difference between the X-ray emissivity-weighted entropy and the entropy inferred from the accretion rate: the accretion routine is measuring the surrounding gas properties in a kernel-weighted average of all gas cells in a spherical shell enclosing $64$ weighted neighbours \citep[see][for details]{Weinberger2023}. This, notably, includes recently launched jets that fill parts of this region, reducing the measured density and increasing the sound speed.\footnote{This aspect is a key feature for obtaining a self-regulated jet power in the simulation.} We can correct for this effect by measuring the average jet volume filling fraction $\alpha$ (defined as the volume occupied by cells with a jet scalar exceeding $10^{-3}$ normalized by the total volume within a radius), while assuming the remainder is filled with gas of cluster electron entropy $K_\mathrm{e,ICM}$,
\begin{align}
    K_\mathrm{e,\,ICM} = K_\mathrm{e} \, (1-\alpha)^{\gamma},
    \label{eq:entropy_icm}
\end{align}
where we assumed that the jet material is in pressure equilibrium with the ICM and its mass is negligible. In the simulation, we measure $\alpha=\left\{0.67, 0.63, 0.85\right\}$ for $M_\bullet=\left\{10^{9}, 10^{10}, 10^{11}\right\}\,\mathrm{M}_\odot\,h^{-1}$. Using Eqs.~\eqref{eq:model_lum} and \eqref{eq:entropy_icm}, and the cooling luminosity in the central $100~\mathrm{kpc}$, we obtain an equilibrium ICM entropy of $K_\mathrm{e,\,ICM}=\left\{0.2,6,41\right\}\,\mathrm{keV}\,\mathrm{cm}^2$, respectively.
 
Figure~\ref{fig:phasediagram} shows a temperature-density phase diagram, colour coded with the mass in each bin for three of the simulations, for conciseness. The solid grey line indicates an entropy of $30~\mathrm{keV}\,\mathrm{cm}^2$, the dashed orange line indicates the respective equilibrium ICM entropy (which is constant along this line and decreases towards the lower right). The phase distribution of the gas shows two important features: (1) The yellow-green narrow band shows the ICM with the low-density part at large radii and higher densities corresponding to smaller radii (as the temperature change in this phase is small). (2) The diagonal blue feature ranging from low-density, very high temperature to higher density, lower temperature is a signature of jets, mixing isobarically with the surroundings. The most relevant aspect for self-regulation is the central entropy value, i.e., where the temperature--density relation is truncated to the right in this diagram as a result of feedback. As is evident in Fig.~\ref{fig:phasediagram}, the simulations differ markedly in this respect: the setup with the lower-mass SMBH exhibits substantially higher central ICM densities, lower central temperatures, and consequently reduced central entropies. Some of the low-entropy material undergoes thermal instability, reaching even higher densities and lower temperatures and eventually forms stars. The derived equilibrium ICM entropies are an excellent lower limit to the entropies reached by most of the gas.

\section{Discussion and Conclusions}
\label{sec:conclusion}

We have derived the central cluster entropy in a cool-core galaxy cluster assuming AGN jet heating--cooling balance in a time-averaged sense, Bondi-like accretion from kpc scales and a fixed  jet feedback efficiency. In particular, we show here that self-regulated systems, which accrete close to the Bondi rate, have a steep scaling with SMBH mass to the point where too massive SMBHs preclude the formation of cool-core clusters because of their efficient feedback. This provides a novel way in providing an upper limit to SMBH masses of $1.4\times10^{10}\,\mathrm{M}_\odot$ in a cool core cluster of mass $10^{15}\,\mathrm{M}_\odot$, assuming a constant jet efficiency of 0.2. This argument complements accretion-based upper limits on SMBH masses \citep{Natarajan2009, King2016}. Such a correlation between central cluster entropy and black hole mass could, however, be weakened by variable efficiencies \citep{GonzalezVillalba2025} caused by a scatter in black hole spins \citep{Sala2024} or environmental dependencies of jet propagation.

We carried out three-dimensional hydrodynamical simulations of an idealized Perseus-like galaxy cluster with AGN jets and find that an increasing SMBH mass implies an elevated central entropy value, for a fixed AGN jet efficiency. We found that successfully reproducing the predictions of our analytic model required corrections for elevated jet entropies when calculating X-ray emissivity-weighted cluster entropies. We note that our subgrid model for SMBH accretion couples to the combined entropy of the ambient ICM and the jet. By construction, this reduces the Bondi accretion rate and, together with the more preventive nature of jet feedback compared to a `kinetic-wind model,' leads to gentler heating of the surrounding ICM \citep{Weinberger2023}, thereby enabling the formation of cool cores with central entropies well below $30\,\mathrm{keV}\,\mathrm{cm}^2$. The ability to displace part of the material in the accretion region while keeping some ICM gas in place is a consequence of details in the numerical choices of the employed jet model and prevents an over-heating to entropies above $30\,\mathrm{keV}\,\mathrm{cm}^2$ in Fig.~\ref{fig:entropyprofile}. Notably, this behaviour is not guaranteed in other feedback models, and heating to non-cool-core states can occur.

Our findings have important implications for cosmological models of galaxy cluster evolution that are simulated with a `kinetic-wind model' \citep{Weinberger2017}. We have shown that high AGN feedback efficiencies in combination with overmassive SMBHs (either via too optimistic accretion histories or via over-merging of massive SMBHs, \citealt{Bahe2022}) lead to central entropies in massive galaxy clusters that exceed the cool-core threshold. We speculate that the absence or rarity in particular of strong, high mass cool-core clusters in some simulation models \citep[e.g.][]{Pakmor2023} can indeed be explained by this effect. Future work will further address the role of cosmological assembly and halo growth on the equilibrium state discussed here. Note that the efficiency parameter in this work, $\epsilon$, represents the product of a possible Bondi-boost factor, (in some models) the radiative efficiency and the feedback efficiency \citep{Weinberger2017,Henden2018}. 

Finally, the analytic scaling of core entropy with SMBH mass presented in this work heavily relies on our assumptions of a Bondi-like accretion. This approach is used to account for the behaviour seen in hydrodynamical simulations, which frequently utilize a version of this formula. Future work will extend this study to alternative accretion models \citep{Weinberger2025} and investigate the physical origin of the relatively small differences reported between chaotic cold accretion and Bondi accretion in galaxy cluster simulations \citep{Meece2017, Ehlert2023}.

\begin{acknowledgements}
RW acknowledges funding of a Leibniz Junior Research Group (project number J131/2022). CP acknowledges support by the European Research Council under ERC-AdG grant PICOGAL-101019746 and from the Deutsche Forschungsgemeinschaft (DFG, German Research Foundation) as part of the DFG Research Unit FOR5195 – project number 443220636.
\end{acknowledgements}

\bibliographystyle{aa}

\end{document}